\documentclass[3p,times,twocolumn]{elsarticle}

\usepackage{ecrc}

\volume{00}

\firstpage{1}

\journalname{Nuclear Physics B Proceedings Supplement}

\runauth{M. Cahill-Rowley}

\jid{nuphbp}

\jnltitlelogo{Nuclear Physics B Proceedings Supplement}

\usepackage{amssymb}
\usepackage[figuresright]{rotating}

\begin{document}

\begin{frontmatter}



\dochead{}

\title{The LHC Confronts the pMSSM}


\author{Matthew Cahill-Rowley}

\address{SLAC National Accelerator Laboratory, Menlo Park, CA, USA\footnote{mrowley@slac.stanford.edu}}

\begin{abstract}
We explore the impact of current (7+8 TeV) and future (14 TeV) LHC searches on the range of viable sparticle spectra within the 19/20 - dimensional phenomenological MSSM (pMSSM). Considering both neutralino and gravitino LSPs, we compare our results with simplified model exclusion limits and describe important cases where the pMSSM results differ significantly from the simplified model descriptions. We also consider models that are poorly constrained by LHC data because of unusual decay topologies and/or displaced decays, and discuss ways to improve the LHC sensitivity in these scenarios. Finally, motivated by naturalness, we examine the sensitivity of current searches to models with light stops and to a specialized set of models with fine-tuning better than 1\%. We show that the 14 TeV LHC will be a very powerful probe of natural pMSSM models.
\end{abstract}


\end{frontmatter}

% \linenumbers

\section{Introduction}
\label{intro}

Although the first run of the LHC was highly successful and culminated in the discovery of a SM-like Higgs boson, it failed to uncover definitive evidence for new physics. While this failure could indicate that the new particles have masses beyond the reach of the LHC, it is also possible that light new particles have simply been hidden by large backgrounds from standard model processes. As a result, a broad experimental and theoretical program has focused on understanding and improving the LHC's sensitivity to a broad range of new physics scenarios. These efforts range from specific studies of UV-complete theories (such as minimal supergravity) to generic studies employing simplified models (in which only one or two particles are considered) or even effective operators. Each technique has important advantages and limitations. Specific theories are highly predictive, but exploring the entire range of possible theories (even qualitatively) is clearly impossible. Simplified models do a poor job of describing cases in which several particles are important for the process of interest, and for considering correlations with \textit{e.g.} dark matter observables. Effective operators are valid only in cases where the intermediate physics is heavy enough to be integrated out, which can be problematic given the large mass scales accessible at the LHC. In this paper, we study the LHC signatures of the minimal supersymmetric standard model (MSSM) using the framework of the pMSSM, which shares some of the advantages of both specific and generic approaches. The results presented here were obtained as part of the Snowmass planning process. Additional details about our methods can be found in \cite{Cahill-Rowley:2014twa} and the references contained therein. 

The MSSM is perhaps the best-motivated theory for physics beyond the Standard Model, due to its ability to solve the hierarchy problem, unify the SM gauge couplings, and potentially provide a stable dark matter candidate. Unfortunately, the general MSSM contains over 100 free parameters. Although we expect that many of these parameters are fixed by details of the UV-complete theory, such as the mechanism of SUSY breaking and possibly the breakdown of a unified gauge group, it is by no means guaranteed that we can anticipate the specific theory realized in nature. In particular, it is possible that a theory with non-minimal particle content could reduce to the MSSM in the low energy limit, suggesting that we should treat the MSSM as an effective theory for which the UV completion is as yet unknown. Fortunately, this approach is aided by very strong experimental constraints on many of the parameters of the general MSSM Lagrangian, motivating an approach in which we apply experimentally-motivated assumptions to simplify the general MSSM. Specifically, we derive the pMSSM from the general R-parity conserving MSSM by assuming ($i$) CP conservation, ($ii$) Minimal Flavor Violation at the electroweak scale so that flavor physics is essentially controlled by the CKM mixing matrix, ($iii$) degenerate 1\textsuperscript{st} and 2\textsuperscript{nd} 
generation sfermion soft mass parameters ({\it e.g.}, right-handed up and charm squarks are degenerate apart from small corrections due to non-zero quark masses), and ($iv$) negligible 
Yukawa couplings and A terms for the first two generations. After these assumptions, the pMSSM is characterized by the 19 parameters listed in Table~\ref{ScanRanges}, with an additional parameter, the gravitino mass, becoming important in the case where the gravitino is the lightest supersymmetric particle (LSP). Although a complete scan of this smaller space is still clearly impossible, a numerical scan is still useful to search for and understand model characteristics which may impede discovery at the LHC. In particular, it is clear that many observables will only depend on a small subset of the parameters, in which case our scan will more fully explore the possible range of phenomenology for that observable.

\begin{table}
\centering
\begin{tabular}{|c|} \hline\hline
$m_{\tilde L(e)_{1/2,3}}$ \\ 
$m_{\tilde Q(u,d)_{1/2}}$ \\ 
$m_{\tilde Q(u,d)_{3}}$  \\
$|M_1|$  \\
$|M_2|$  \\
$|\mu|$ \\ 
$M_3$ \\ 
$|A_{t,b,\tau}|$  \\ 
$M_A$  \\ 
$\tan \beta$ \\
\hline\hline
\end{tabular}
\caption{The 19 parameters of the pMSSM with a neutralino LSP. Models with a gravitino LSP are described by the same 19 parameters plus the gravitino mass $m(\tilde{G})$.}
\label{ScanRanges}
\end{table}

\section{Methods}
\label{methods}

In this paper, we study the LHC phenomenology of several sets of pMSSM models that are consistent with the set of pre-LHC experimental data and theoretically viable. In particular, we study $\sim$223k models with a neutralino LSP and $\sim$229k models with a gravitino LSP, all of which predict a LSP abundance that is \textit{less than or equal to} the total dark matter abundance. The specific constraints applied to select these models are detailed in \cite{CahillRowley:2012cb}. We also study a special set of $\sim$10.2k models with a neutralino LSP that have low fine tuning ($\Delta < 100$) according to the Barbieri-Giudice measure and additionally predict an LSP abundance \textit{equal to} the total dark matter abundance, the generation of which was described in \cite{Cahill-Rowley:2014twa}. Although many of the models in the general neutralino and gravitino model sets do not predict the observed Higgs mass, we have found that the LHC searches are approximately independent of the Higgs mass requirement within the allowed range of masses. In order to obtain a reasonably comprehensive picture of the LHC SUSY results, we simulate the ATLAS and CMS searches listed in Tables~\ref{SearchList7} and~\ref{SearchList8} and determine whether each model is excluded or allowed by the combined search results. We also simulate the potential impact of null results from ATLAS Jets+MET and stop searches at 14 TeV with both 300 and 3000 fb$^{-1}$. Due to the large amount of CPU resources required, we simulate 14 TeV events only for the subsets of models predicting the correct Higgs mass. For all of these simulations, we generate events with PYTHIA 6.4.26 \cite{Sjostrand:2006za} and scale the event rates to NLO using K-factors calculated by Prospino 2.1 \cite{Beenakker:1996ch}. We then employ PGS 4~\cite{PGS} for our fast detector simulation. Both PYTHIA and PGS have been modified to correctly deal with gravitinos, multi-body decays, hadronization of stable colored sparticles, and ATLAS b-tagging. Finally, we apply the cuts for the simulated analyses using our customized analysis code, which follows the cuts and selection criteria employed in the searches as closely as possible. Models are considered excluded if the event rate in any signal region is above the 95\% CL$_s$ limit set by ATLAS or CMS. We validate our results by running our simulation on benchmark models for each signal region in the various analyses and comparing our results with the published selection efficiencies. 

\begin{table}[htb]
\centering
\begin{tabular}{|l|l|} \hline\hline
Search & Reference   \\
\hline
2-6 jets & ATLAS-CONF-2012-033   \\
multijets & ATLAS-CONF-2012-037 \\
1 lepton & ATLAS-CONF-2012-041  \\

HSCP      &  1205.0272   \\
Disappearing Track  & ATLAS-CONF-2012-111 \\
Muon + Displaced Vertex  & 1210.7451  \\
Displaced Dilepton & 1211.2472  \\

Gluino $\to$ Stop/Sbottom   & 1207.4686 \\
Very Light Stop  & ATLAS-CONF-2012-059    \\
Medium Stop  & ATLAS-CONF-2012-071 \\
Heavy Stop (0$\ell$)  & 1208.1447  \\
Heavy Stop (1$\ell$)   & 1208.2590  \\
GMSB Direct Stop  & 1204.6736 \\
Direct Sbottom & ATLAS-CONF-2012-106 \\
3 leptons & ATLAS-CONF-2012-108  \\
1-2 leptons & 1208.4688  \\
Direct slepton/gaugino (2$\ell$)  & 1208.2884 \\
Direct gaugino (3$\ell$) & 1208.3144 \\
4 leptons & 1210.4457  \\
1 lepton + many jets & ATLAS-CONF-2012-140  \\
1 lepton + $\gamma$ & ATLAS-CONF-2012-144  \\
$\gamma$ + b & 1211.1167  \\
$\gamma \gamma $ + MET & 1209.0753  \\

$B_s \to \mu \mu$ & 1211.2674 \\
$A/H \to \tau \tau$ & CMS-PAS-HIG-12-050 \\

\hline\hline
\end{tabular}
\caption{7 TeV LHC searches included in the present analysis.}
\label{SearchList7}
\end{table}

\begin{table}[hbt]
\centering
\begin{tabular}{|l|l|} \hline\hline
Search & Reference   \\
\hline

2-6 jets   & ATLAS-CONF-2012-109  \\
multijets   & ATLAS-CONF-2012-103  \\
1 lepton     & ATLAS-CONF-2012-104  \\
SS dileptons & ATLAS-CONF-2012-105  \\
2-6 jets   & ATLAS-CONF-2013-047  \\

HSCP      &  1305.0491  \\

Medium Stop (2$\ell$) & ATLAS-CONF-2012-167  \\
Medium/Heavy Stop (1$\ell$) & ATLAS-CONF-2012-166  \\
Direct Sbottom (2b) & ATLAS-CONF-2012-165  \\
3rd Generation Squarks (3b) & ATLAS-CONF-2012-145 \\
3rd Generation Squarks (3$\ell$) & ATLAS-CONF-2012-151  \\
3 leptons & ATLAS-CONF-2012-154  \\
4 leptons & ATLAS-CONF-2012-153  \\
Z + jets + MET & ATLAS-CONF-2012-152  \\

\hline\hline
\end{tabular}
\caption{8 TeV searches included in this analysis. The combination of searches in this table and the previous table exclude $\sim 45.5~(61.3,~74.0)\%$ of models in the the neutralino (gravitino, low-FT) model sets.}
\label{SearchList8}
\end{table}

\section{Results}
\label{results} 

\subsection{Neutralino LSP}
\label{neut}
\begin{figure}[htb]
\centerline{\includegraphics[width=2.7in]{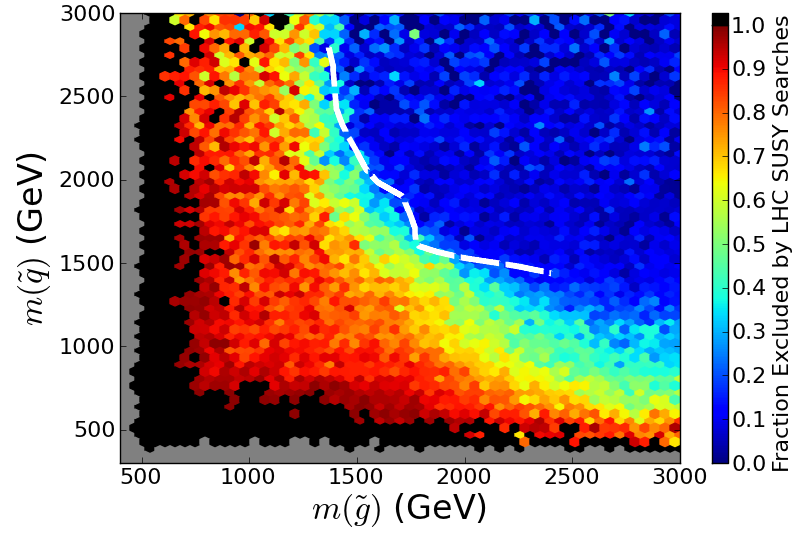}}
\centerline{\includegraphics[width=2.7in]{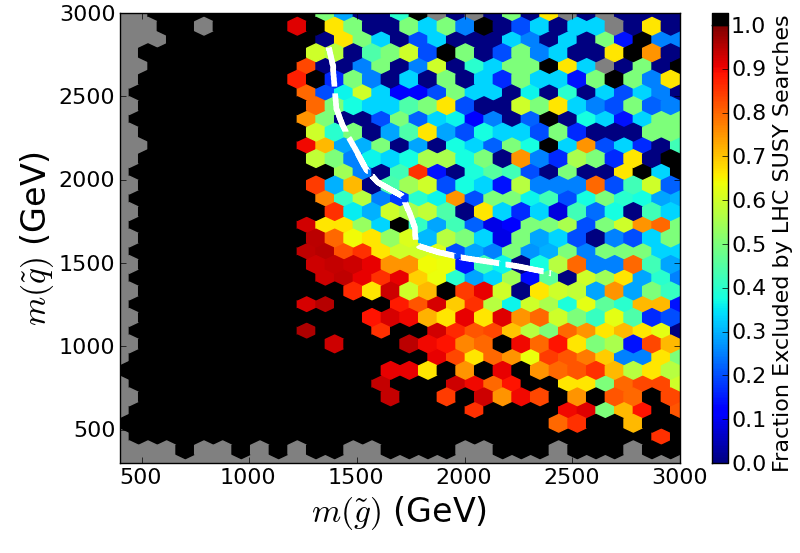}}
\vspace*{-0.30cm}
\caption{The fraction of models excluded by the combined LHC searches, projected in the lightest squark-gluino mass plane, for the general neutralino (top) and Low-FT (bottom) model sets. In this and all subsequent figures, solid or dashed white lines indicate 8 TeV simplified model limits.}
\label{neut:squark-gluino}
\end{figure}

We first examine models with a neutralino LSP, including both the general neutralino and Low-FT model sets described above. Since the LHC is a hadron collider, we expect large production rates for colored sparticles, particularly 1$^{st}$/2$^{nd}$ generation squarks and gluinos, and correspondingly strong limits on their masses. Figure~\ref{neut:squark-gluino} shows 
that this is in fact the case - most models with squarks lighter than $\sim$ 800 GeV or gluinos lighter than $\sim$ 1.4 TeV are excluded by the LHC searches. On the other hand, the exclusion fraction is quite low for models with heavy squarks and gluinos, indicating that direct searches for 3$^{rd}$ generation squarks and color singlet sparticles are currently sensitive to a comparatively small portion of the parameter space. Surviving models with light squarks and gluinos have relatively heavy LSP masses, producing compressed spectra which are difficult to see given the large QCD backgrounds. The lower panel of Figure~\ref{neut:squark-gluino} shows that this effect is mostly avoided in the Low-FT model set, since the fine-tuning requirements necessitate a LSP lighter than $\sim$ 400 GeV, dramatically increasing the effectiveness of LHC searches for Low-FT models. While the presence of non-standard decay pathways can also degrade the exclusion limits, this effect is typically sub-dominant to the dramatic effects of spectrum compression. 

\begin{figure}[hbt]
\centerline{\includegraphics[width=2.7in]{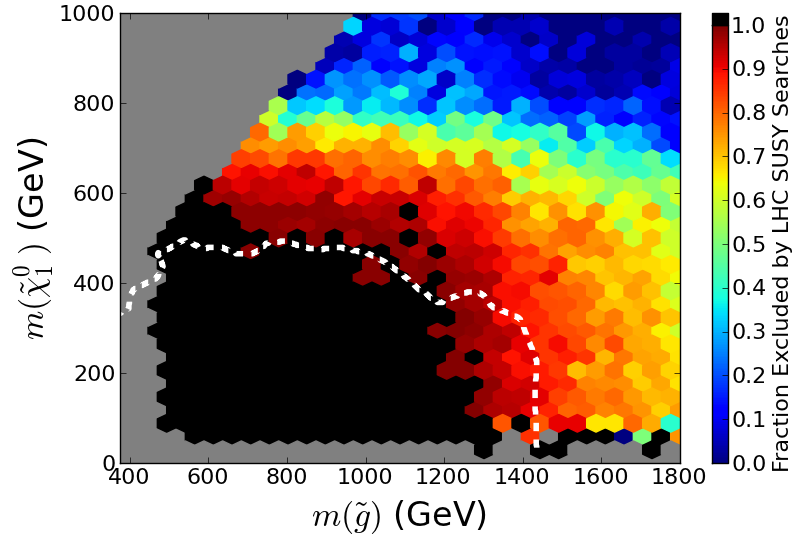}}
\centerline{\includegraphics[width=2.7in]{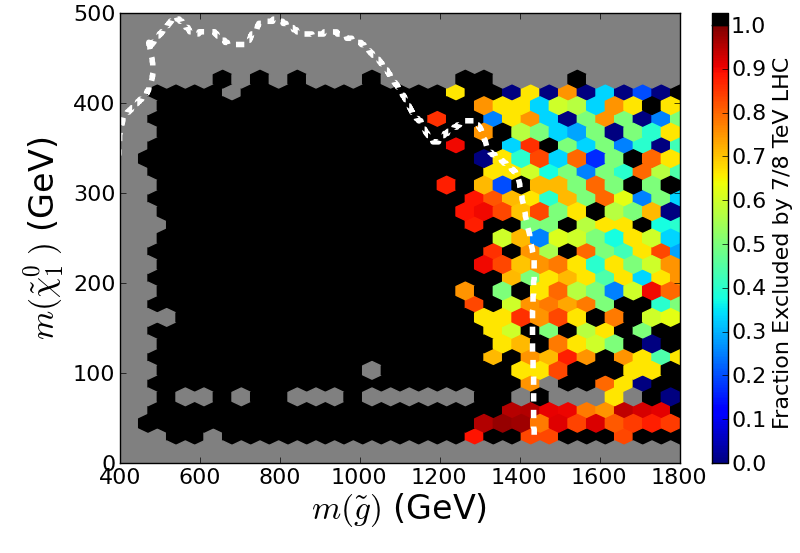}}
\vspace*{-0.30cm}
\caption{The fraction of models excluded by the combined LHC searches, projected in the LSP-gluino mass plane, for the general neutralino (top) and Low-FT (bottom) model sets.}
\label{neut:gluino}
\end{figure}

\begin{figure}[htb]
\centerline{\includegraphics[width=2.7in]{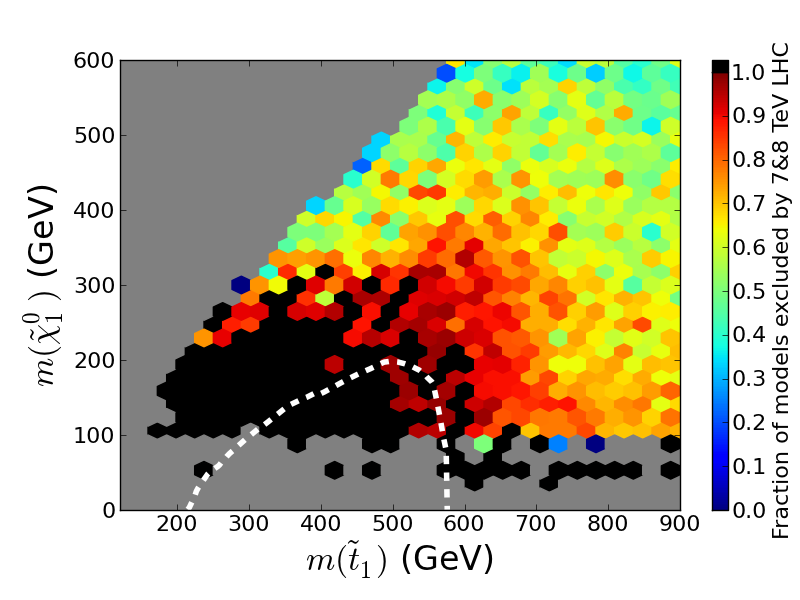}}

\centerline{\includegraphics[width=2.7in]{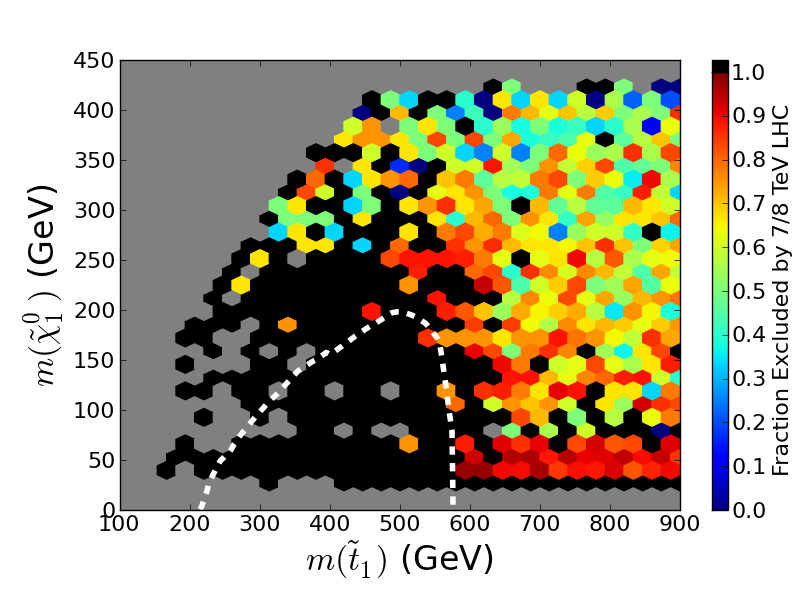}}
\vspace*{-0.30cm}
\caption{The fraction of models excluded by the combined LHC searches, projected in the LSP-light stop mass plane, for the general neutralino (top) and Low-FT (bottom) model sets.}
\label{neut:stop}
\end{figure}

We now turn our attention to the limits on stops and gluinos, both of which are important for naturalness. Figure~\ref{neut:gluino} shows the gluino-LSP mass plane for both model sets. In both cases, we see that spectrum compression is mainly responsible for determining the shape of the excluded region and is well-described by the simplified model limit. In both model sets, some models survive below the simplified model limit for large gluino masses, above $\sim$ 1.2 TeV. In these models, the gluino decays typically produce 3$^{rd}$ generation quarks, in which case the expected limit comes mainly from a search for final states containing 3 b-jets and is slightly weaker. The bottom panel of the figure shows the importance of the upper limit on the LSP mass in the low-FT model set, which has effectively doubled the minimum viable gluino mass from $\sim$600 GeV to $\sim$1.2 TeV. Figure~\ref{neut:stop} shows the stop-LSP mass plane. In this case, we see that the effect of spectrum compression is not nearly as severe as for the simplified model limit (which assumes a 100\% branching fraction to the $t \chi^0$ final state). The reason for this difference is that most of our models have a relatively small splitting between the lightest chargino and the LSP{\footnote{Most models in the general neutralino model set have wino-like or Higgsino-like LSPs, since bino-like LSPs require a specific annihilation mechanism to avoid overclosing the universe. Even when the LSP is bino-like, as it is in the low-FT model set, a sizable Higgsino component is frequently required to boost the annihilation rate, and the charged Higgsino is typically within $\sim$ 40 GeV of the LSP mass.}}. This means that stop decays to $b \chi^-$ final states are nearly always allowed, and dominate in the compressed spectrum region. This decay mode produces relatively hard b-jets, which are effectively targeted by the direct sbottom search, particularly when the chargino decay products are very soft. In particular, the direct sbottom search has dedicated signal regions which make use of initial state radiation to effectively probe the compressed spectrum region, resulting in the strong exclusion limits shown in Figure~\ref{neut:stop}.


\subsection{Gravitino LSP}
\label{grav}

We can perform a similar analysis on models with a gravitino LSP. The main difference with respect to the neutralino LSP scenario results from the Planck-suppressed coupling between the gravitino and other sparticles, particularly the next-to-lightest supersymmetric particle (NLSP), which is forced by R-parity conservation to decay to the gravitino with a width given by:

\begin{equation}
\Gamma_{NLSP} \sim \frac{m_{NLSP}^5}{48 \pi M_{Planck}^2 m_{\tilde{G}}^2}
\end{equation}

\begin{figure}[hbt]
\centerline{\includegraphics[width=2.7in]{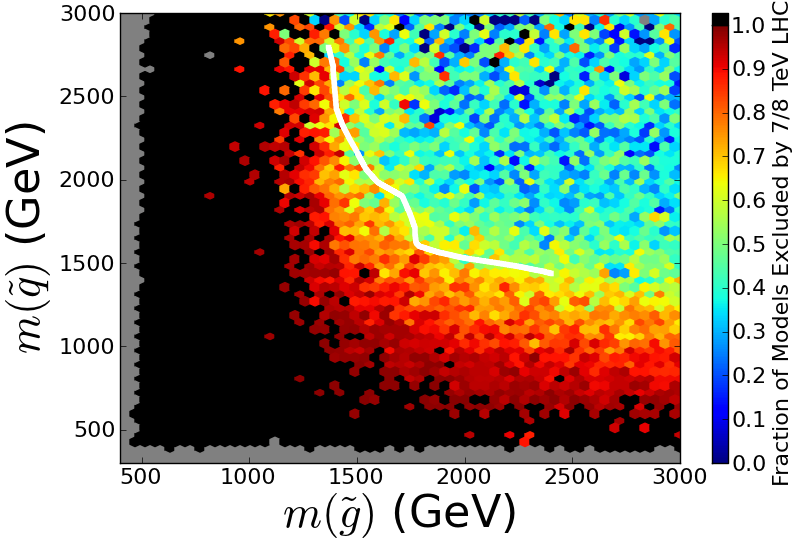}}

\centerline{\includegraphics[width=2.7in]{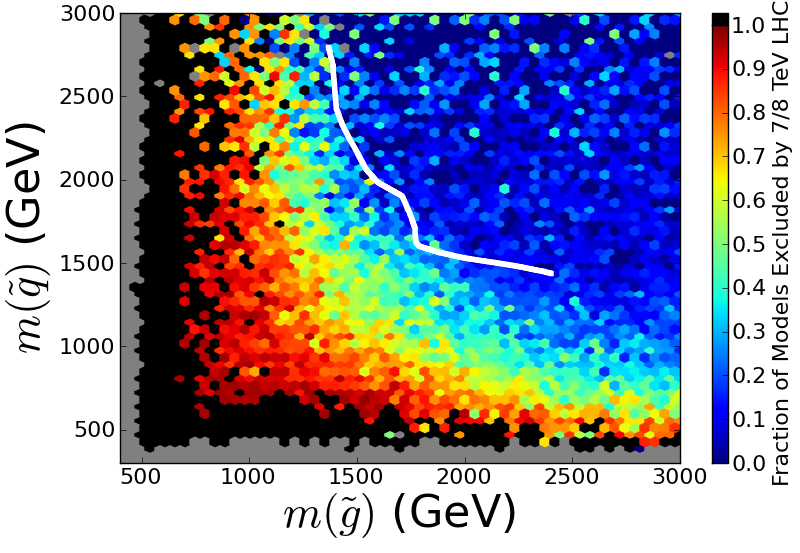}}
\vspace*{-0.30cm}
\caption{Same as Figure 1, but now for visible (upper panel) and invisible (lower panel) NLSPs in the gravitino model set.}
\label{grav:squark-gluino}
\end{figure}

Depending on the gravitino and NLSP masses, the NLSP can decay promptly, within the detector, or even well outside of the detector. Our model sample contains many examples of all 3 scenarios. Since the NLSP width decreases with increasing gravitino mass, prompt NLSP decays require the gravitino mass to be less than $\sim$1 KeV. Models with a gravitino heavy enough to produce a true compressed spectrum therefore have extremely long-lived NLSPs. However, if the NLSP is a stable neutralino or a sneutrino, it will produce only missing energy and act as the LSP for the purposes of collider experiments, potentially resulting in an effectively compressed spectrum even when the gravitino is light. Since spectrum compression was so important in our neutralino LSP results, we analyze these ``invisible NLSP'' models separately from the remaining models, which we will refer to as having ``visible NLSPs''. Figure~\ref{grav:squark-gluino} shows the lightest squark-gluino mass plane for the gravitino model set after making this division. Comparing with Figure~\ref{neut:squark-gluino}, we see that the limit on models with visible NLSPs is substantially improved over the neutralino LSP case. Additionally, the LHC searches are sensitive to a far larger percentage of models with negligible squark and gluino production. This is partly due to the prevalence of detector-stable charged NLSPs, which are subject to strong limits from searches for heavy stable charged particles, and partly a result of the improved effectiveness of searches for electroweakinos and sleptons when the LSP mass is very small. On the other hand, the lower panel (showing models with invisible NLSPs) looks very similar to the analogous result for the general neutralino model set in Figure~\ref{neut:squark-gluino}, although we see that the reach is actually slightly worse for invisible NLSP models. In particular, as we will see below for specific examples, models with sneutrino NLSPs are less constrained by LHC searches. Finally, we note that even in the visible NLSP scenario, some models survive significantly below the simplified model limit. Although some of these models simply have unusual decay topologies involving \textit{e.g.} tau leptons, many of them feature displaced NLSP decays to leptons or jets. In this case, signal events are frequently rejected by quality cuts, including a requirement that jets should contain a certain fraction of charged tracks pointing to the primary vertex, a veto on muons with a large impact parameter (to reject cosmic rays), and the requirement that an electron candidate must be associated with a charged track that has a small impact parameter. While some dedicated searches for displaced objects exist, and several are included in Table~\ref{SearchList7}, they target different types of models and are insensitive to many of the NLSP decay modes found in our model set.  

\begin{figure}[htb]
\centerline{\includegraphics[width=2.7in]{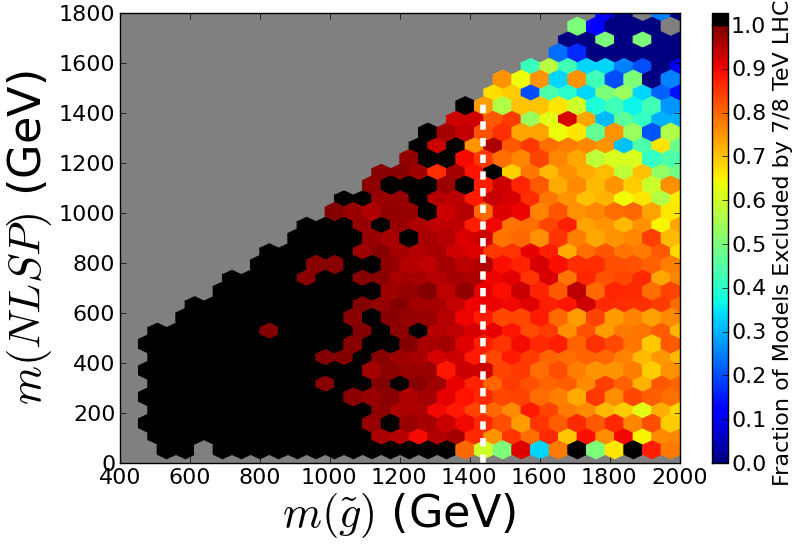}}

\centerline{\includegraphics[width=2.7in]{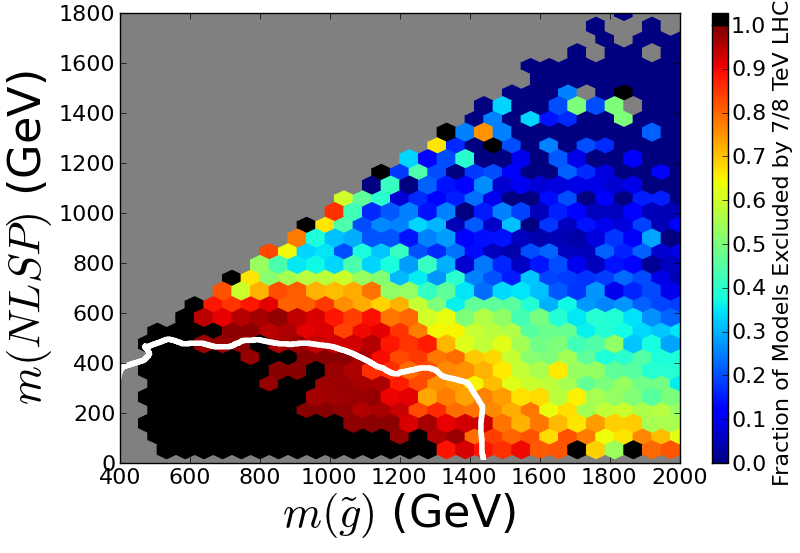}}
\vspace*{-0.30cm}
\caption{Same as Figure 2, but now for visible (upper panel) and invisible (lower panel) NLSPs in the gravitino model set.}
\label{grav:gluino}
\end{figure}

It is again interesting to specifically examine the search reach for models with light stops and gluinos. Figure~\ref{grav:gluino} shows the fraction of models excluded in the gluino-NLSP mass plane, divided by NLSP visibility. In the top panel (visible NLSPs), the excluded fraction is roughly independent of the LSP mass, as we might expect given that hard decay products can result either from gluino decays to the NLSP or from NLSP decays to the gravitino. However, models do survive well below the limit that we would expect for a gluino decaying directly to a massless LSP. Interestingly, all of the surviving models with gluinos lighter than $\sim$ 1.1 TeV have stau NLSPs with macroscopic decay lengths; models with slightly heavier gluinos can also have other NLSPs with macroscopic decay lengths or a promptly decaying stau NLSP. It would be interesting to see whether searches specifically targeting tau final states are sensitive to the latter scenario, although this would require an accurate simulation of the tau tagging efficiency which is not possible using PGS. The bottom panel of Figure~\ref{grav:gluino} shows invisible NLSP models. In this case, the models surviving below the simplified model limit all have gluinos decaying to sneutrino NLSPs, except for gluino masses above 1.2 TeV, where models with stable neutralinos are also seen to survive in Figure~\ref{neut:gluino}. In models with a sneutrino NLSP, each gluino decay results in a minimum of 4 decay products, which frequently include relatively soft leptons. While these soft decay products are unlikely to be seen in generic searches, the high multiplicity of final state particles could allow a targeted search to significantly improve the LHC sensitivity to these models.

\begin{figure}[hbt]
\centerline{\includegraphics[width=2.7in]{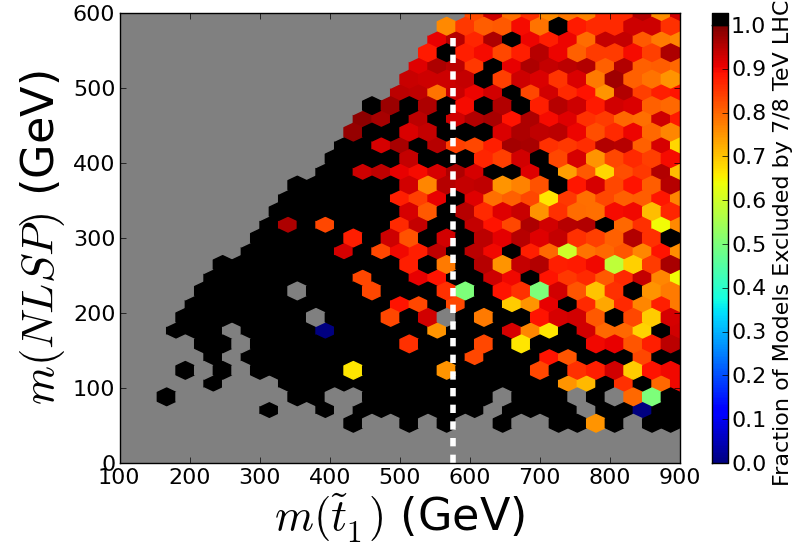}}
\centerline{\includegraphics[width=2.7in]{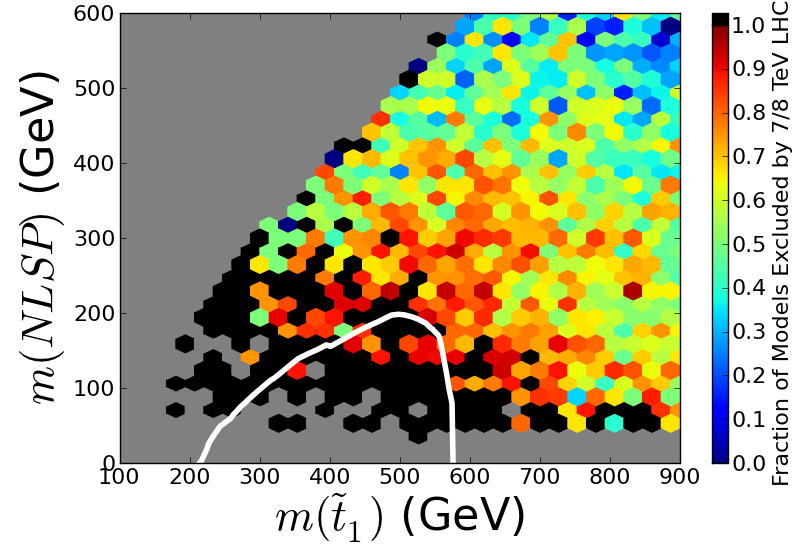}}
\vspace*{-0.30cm}
\caption{Same as Figure 3, but now for visible (upper panel) and invisible (lower panel) NLSPs in the gravitino model set.}
\label{grav:stop}
\end{figure}

Interestingly, exactly the same pattern is found to hold for models with light stops, shown in Figure~\ref{grav:stop}. Specifically, models with light stops and visible NLSPs generally have displaced and/or stau NLSPs. Similarly, models with invisible NLSPs that survive below the simplified model limit have stops decaying to sneutrino NLSPs. This similarity to the gluino case is surprising given that very different searches are providing the majority of the sensitivity in each case, and showcases the much greater role played by the NLSP in models with a gravitino NLSP. In particular, if the NLSP is close to the mass of the produced sparticle, only the NLSP decay products are hard enough to be seen, and identity of the produced sparticle is only relevant for determining the rate of NLSP production! 

\begin{figure}[hbt]
\centerline{\includegraphics[width=2.7in]{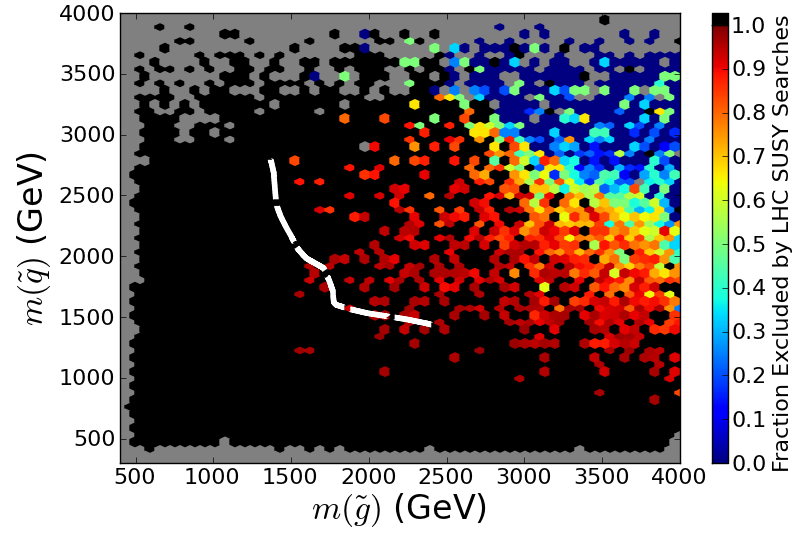}}
\centerline{\includegraphics[width=2.7in]{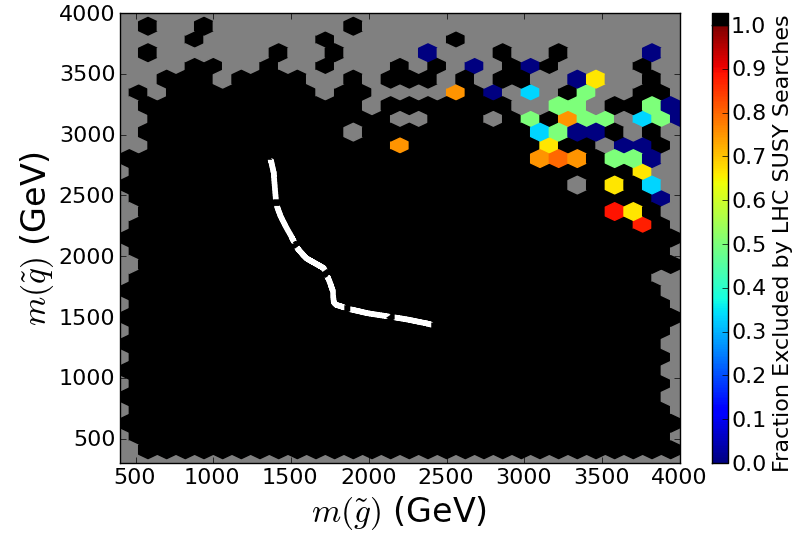}}
\centerline{\includegraphics[width=2.7in]{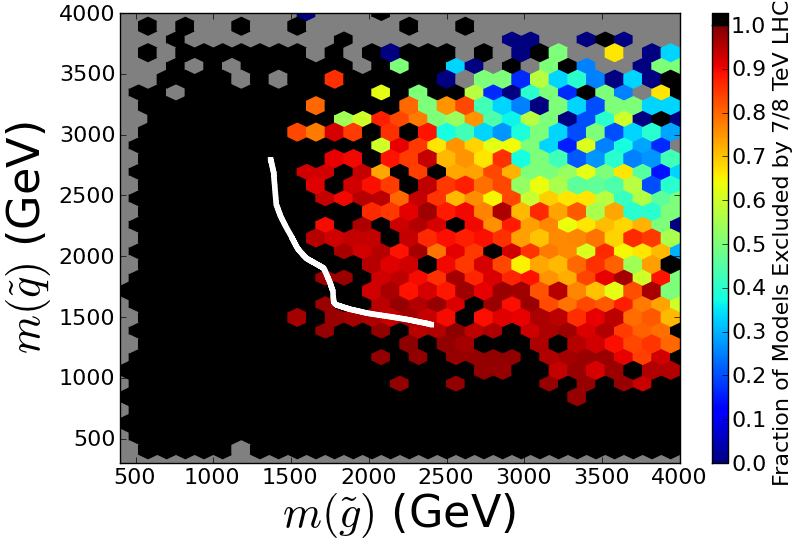}}
\vspace*{-0.30cm}
\caption{Fraction of models excluded by the combined 7,8, and 14 TeV searches in the lightest squark-gluino mass plane. The top, middle, and bottom panels show the neutralino, gravitino, and low-FT model sets, respectively. As described above, only models with the correct Higgs mass are shown in this figure.}
\label{14-300}
\end{figure}

Finally, we can examine the prospects for sparticle searches at 14 TeV. Figure~\ref{14-300} shows the expected results for each model set with 300 fb$^{-1}$ of integrated luminosity in the squark-gluino mass plane. It is important to remember that these limits result from only 3 searches (Jets+MET, 0l and 1l stop searches), and that there are therefore many cases in which the searches are poorly matched to the produced final states, particularly for gravitino LSP models. However, we can see the sharp increase in the reach for models to which the simulated searches are sensitive. Once again the sensitivity to the low-FT model set is very high. Indeed, going to 3 ab$^{-1}$ we find that all of the low-FT models are expected to be excluded by this combination of searches.

\section{Conclusions}

From our survey of these different pMSSM model sets, we have seen that the LHC searches have a robust sensitivity to a wide range of SUSY models and decay topologies. This sensitivity is particularly impressive in the case of models with gravitino LSPs, since many of the decay modes have not been the subject of targeted searches. We have found that searches for colored sparticles decaying to staus and sneutrinos, as well as additional searches targeting displaced decays to jets and leptons, could substantially improve the coverage of the gravitino LSP scenario. However, we have also seen that spectrum compression is the dominant mechanism allowing light sparticles to remain undetected, validating the large experimental effort that has gone into improving sensitivity to the compressed regions. A particularly good example of this is the exploitation of ISR in the direct sbottom search, which improves the sensitivity of the search to a large class of models. We expect that the increased energy and luminosity in the LHC run 2, combined with continued improvement of the analysis techniques, will provide a very broad sensitivity to TeV-scale SUSY, particularly models with low fine-tuning and the correct dark matter abundance, hopefully culminating in the discovery of supersymmetry! 

%



%% References

\nocite{*}
\bibliographystyle{elsarticle-num}
\bibliography{references}



%\begin{thebibliography}{00}


%\cite{PGS}
%\bibitem{PGS}
%  J. Conway, PGS4, Pretty Good detector Simulation, http://www.physics.ucdavis.edu/~conway/research/software/pgs/pgs.html.

%\end{thebibliography}

\end{document}